# Experimental study of transport properties of Weyl semimetal LaAlGe thin films grown by molecular beam epitaxy

**Niraj Bhattarai[1,2,*], Andrew W. Forbes[1,2], Christopher Gassen[1,2], Raghad S. H. Saqat[1,2], Ian L. Pegg[1,2], and John Philip[1,2]**

[1]Department of Physics, The Catholic University of America, Washington, DC, 20064, USA.
[2]The Vitreous State Laboratory, The Catholic University of America, Washington, DC, 20064, USA.
*Corresponding author: bhattarai@cua.edu

## ABSTRACT

Rare earth compounds display diverse electronic, magnetic, and magneto-transport properties. Recently these compounds of the type $R$AlGe ($R$ = La, Ce, Pr) have been shown to exhibit Weyl semimetallic behavior. In this work, we have investigated the crystal structure, electronic, and magneto-transport properties of the Weyl semimetal LaAlGe thin films grown by molecular beam epitaxy. The temperature dependence of longitudinal resistivity at different magnetic fields is discussed. Observations of magnetoresistances and Hall effect at different temperatures and their evolution with magnetic field up to 6 T are also discussed with relevant mechanisms. We have observed positive unsaturated magnetoresistances, with a small quadratic contribution at low temperatures, which tends to saturate at higher fields. The Hall measurements confirm the electron-dominated semimetallic conduction with an average charge carrier density of ~ $9.68 \times 10^{21}$ $cm^{-3}$ at room temperature.

## I. INTRODUCTION

Weyl semimetals (WSMs), materials hosting Weyl fermions in the form of low energy electronic excitations that can be described by the Weyl equation, are an interesting class of topological matter that has generated substantial research interest in the present decade[1–5]. The origin of WSM behavior in materials lies in breaking crystal inversion symmetry, time-reversal symmetry, or both [6]. Their identification as WSMs effectively depends on the observation of a few hallmark features in their electronic structure, such as linear dispersion of Weyl fermions through band crossings (Weyl points) and unclosed Fermi-arcs in the surface states[7–9]. The WSMs have been a great platform for several intriguing topological phenomena arising from the interplay of Weyl fermions with electrical and magnetic fields. For example, the Weyl fermions in the fermi arc are predicted to show novel quantum interference effects in tunneling spectroscopy and quantum oscillation in magnetotransport [10–12], whereas the Weyl fermions in the bulk states can give rise to negative magnetoresistance due to non-conservation of chiral charge in magnetic fields and the anomalous Hall effect [6,13–15].

For electronic applications, it is desirable to realize the materials in thin films. The reduced dimensionality allows the control of the topological properties and can lead to interesting physics that may not be explored in their bulk counterparts[16,17]. Recent theoretical observations have shown that thin film WSMs possess promising characteristics such as a thickness-dependent metal-insulator transition[18], twisting Fermi surface[19], spin current without a simultaneous charge current[20], and Floquet topological insulator phases[21]. Also, many experiments have demonstrated that the low-dimensional WSMs can have technological applications in a photovoltaic system[22], mid-infra-red photodetector[23], and quantum computers[24].

LaAlGe is a rare earth based nonmagnetic ternary alloy compound. Recently, first-principles band structure calculations and angle resolved photo emission spectroscopy measurements have confirmed that LaAlGe hosts Lorentz violating type II Weyl nodes generated by breaking of space inversion symmetry [8,25]. It has the LaPtSi prototype crystal structure, where the Pt and Si sites are taken by Al and Ge, and belongs to a space group $I4_1md$ (109) forming a body-centered tetragonal Bravais lattice with parameters a = 4.32 Å = b, and c = 14.80 Å[26,27]. Single crystal of LaAlGe grown by high-temperature self-flux method, show electronic transport dominated by Weyl nodes with tiny effective masses ($\sim 0.024m_0$) and high Fermi velocities ( $\sim 2.77 \times 10^6$ m/s)[28]. Relative to the potential it offers for various modern device applications, this material is quite understudied to explore transport behavior. Only a few published reports on LaAlGe demonstrating some preliminary physical properties such as longitudinal resistivity variation with temperature exist and it largely remains to be explored. Single crystals of LaAlGe display metallic resistivity, small residual resistivity ratio (RRR) ~ 2 with a residual resistivity of about 30 μΩ cm [29]. Another separate study have shown a slightly lower residual resistivity ~ 10 μΩ cm [28]. In this article, we have grown ~50 nm thick films of LaAlGe by molecular beam epitaxy (MBE) and report a systematic and detailed investigation on various electronic and magneto-transport measurements. Interestingly, in the low-temperature regime we observed a slight up-turn in the longitudinal resistivity measurements at magnetic fields higher than 2 T hinting at a possible bandgap opening. Our findings through various experimental investigations confirms the semimetallic nature of LaAlGe films.

## II.    EXPERIMENTAL DETAILS

We have grown 50 nm thick LaAlGe film samples on polished insulating silicon (100) substrates by MBE with a base pressure better than $1.1 \times 10^{-9}$ Torr during deposition and less than $1.1 \times 10^{-8}$ during annealing. The films were annealed at 793 K for 16 hours *in situ* to get a high-quality polycrystalline LaAlGe films. Before deposition, silicon substrates were etched in 20 % hydrofluoric acid solution (removes native oxide) and rinsed with acetone, isopropyl alcohol, and deionized water. The silicon substrates were preheated at 523 K for 30 min and a 10 nm buffer layer of MgO was deposited and annealed at 883 K for one hour [30]. The lanthanum, aluminum and, germanium metals were deposited keeping the substrate temperature at 623 K throughout the deposition. In the end, 5 nm MgO was deposited as a capping layer to prevent the film from further oxidization and the samples were transferred out of the chamber. The detailed growth of LaAlGe thin film used in this study has been previously reported elsewhere[26]. The crystal structure of the film were characterized by X-ray diffraction (XRD), composition by energy dispersive X-ray (EDX) analysis and surface topography by atomic force microscopy (AFM) and scanning electron microscopy (SEM). The longitudinal resistivity $\rho_{xx}$, the magnetic field dependence of longitudinal resistivity, magnetoresistance, and Hall resistivity $\rho_{xy}$ were measured in Quantum Design Physical Properties Measurement System under magnetic fields up to 7 T and temperatures between 2.5 and 300 K. All the electrical measurements were carried out using the AC transport horizontal rotator with conventional four-probe geometry except for Hall effect measurements, which was carried using Hall bar devices with dimension 6 × 6 mm and 50 nm thickness grown by using metal contact masks. Electrical contacts were made with gold wire electrodes attached using indium metal.

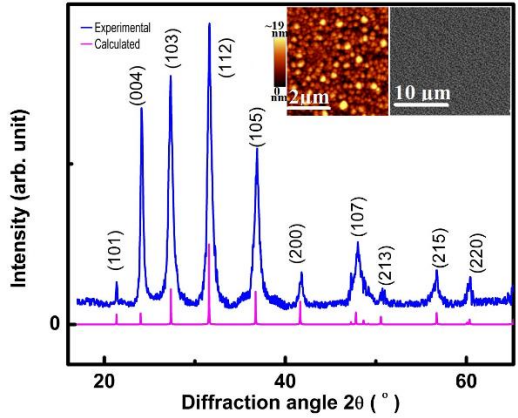

**Figure 1.** X-ray diffraction peaks of the MBE grown 50 nm thick LaAlGe film. The inset in the upper right corner display the SEM and AFM images of the film.

## III. RESULTS AND DISCUSSION

Figure 1 shows the X-ray powder diffraction peaks of 50 nm LaAlGe film grown on a Si(100) substrate. It exhibits a body-centered tetragonal crystal structure with a non-centrosymmetric space group-$I4_1md$ (109). The inset in the upper right corner displays the SEM and AFM images, clearly showing a continuous and uniform surface topography of the film. The EDX analysis was conducted on different batches of LaAlGe film samples. A typical EDX spectrum of a film with La:Al:Ge in the stoichiometric ratio of 33.8:32.3:33.9 is shown in Figure 2. This is in close approximation to the expected 1:1:1 stoichiometric ratio. The longitudinal resistivity $\rho_{xx}$ as a function of temperature ranging from 2.5 to 300 K and under zero magnetic field is shown in Figure 3, which is metal-like behavior. The minimum resistivity observed at 2.5 K is 85.8 μΩ cm, which is slightly higher than was reported for single crystal [29]. The MBE grown thin film sample displays a residual resistivity ratio (RRR) $\rho_{xx}(300\ \text{K})/\rho_{xx}(2.50\ \text{K}) = 1.18(1)$. This compares well with the RRR (=2) value achieved for the best single crystals LaAlGe [28]; and also with the single crystals of its isostructural family members CeAlGe and PrAlGe. The single crystals of $R$AlGe grown by different techniques have shown RRR in the range 1.3 to 2.3 with diverse residual resistivity $\rho_0$ starting from 30 to 228 μΩ cm [1,29,31,32]. This indicates that a small RRR value is an intrinsic property of the $R$AlGe family. The red curve within the temperature range from 2.5 to 42 K represents a fit to $\rho_{xx}$. The form of the fitted equation is $\rho_{xx} = \rho_0 + aT^b$, which is the low temperature approximation of Bloch-Grüneisen formula, where $\rho_0$, $a$, and $b$ are fitting parameters with values 86(1) μΩ cm, 1.09(2) × 10⁻⁵ μΩ cm K⁻³ and 3.05(1) respectively. The term $\rho_0$ is the residual resistivity, which arises due to the impurity scattering. $b \sim 3$ shows there is a deviation from both, pure electronic correlation-dominated scattering mechanism ($b = 2$) according to Fermi-liquid theory, which is observed in WTe$_2$[33], and conventional electron-phonon scattering process ($b = 5$) according to Bloch-Grüneisen theory[34]. In particular, $b$ ranging between 2 and 5 is attributed to a dominant $s$-$d$ interband scattering, which has also been observed in single crystals of LaAlGe[28] and PrAlGe[32] and many other unconventional semimetals including LaSb ($b = 4$) and LaBi ($b = 3$) and a ternary Dirac semimetal ZrSiS ($b = 3$)[35–38].

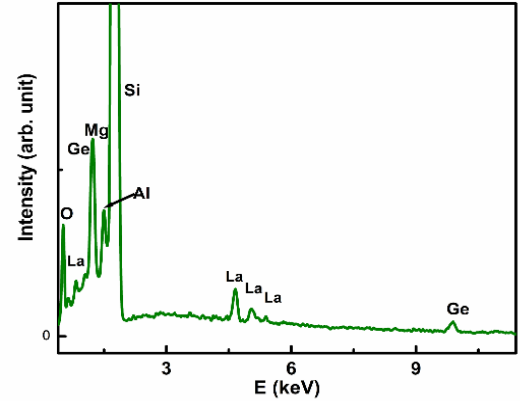

**Figure 2.** Energy dispersive X-ray spectrum of LaAlGe thin film.

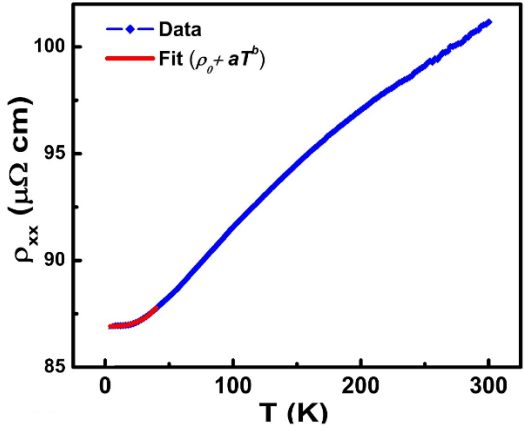

**Figure 3.** Longitudinal resistivity $\rho_{xx}$ as a function of temperature and under zero magnetic field. The red curve represents a fit $\rho_{xx} = \rho_0 + aT^b$ for 2.5 ≤ T (K) ≤ 42.

Figures 4 (a) and (b), respectively, display the $\rho_{xx}$, as a function of temperature and various magnetic fields up to 7 T applied perpendicular and parallel to the direction of current flow. With the application of the magnetic field, the resistivity of the sample imitates the zero-field curve until a critical field is reached, where the $\rho_{xx}$ curve passes through a minimum and shows a noticeable upturn toward lower temperatures. The position of minima shifted to higher temperature under higher magnetic fields. The minimum $\rho_{xx}$ is recorded at a field-dependent temperature T*, also called "turn-on" temperature (taken at the minimum in the resistivity curve). When the field is applied perpendicular to the direction of current flow, the turn-on temperature T* shifted from ~11.70(3) K in a magnetic field of 2 T to ~17.94(3) K in 7 T, which is shown as an inset in Figure 4 (a). On the other hand, measurements are taken with a magnetic field parallel to the direction of current flow resulted in comparatively low turn-on temperatures in comparison to perpendicular field measurements. In parallel 7 T field, T* ~15.76(3) K (the inset in Figure 4 (b)), which is 2 K short to perpendicular measurement at similar field condition. However, in both cases, the T* shifted up with an increasing magnetic fields and tends to stabilize as larger fields are applied.

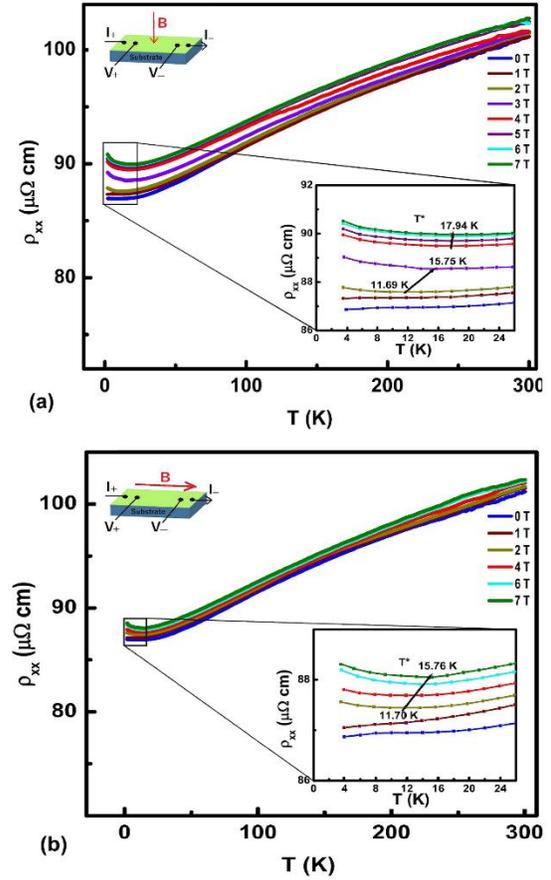

**Figure 4:** Temperature dependence of longitudinal resistivity in some of the selected magnetic fields in different orientations. (a) Magnetic field applied perpendicular to the direction of current flow. (b) Magnetic field applied parallel to the direction of current flow. The schematics for the direction of magnetic field and current is shown in the upper left corner of both figures. Insets in the lower right corner display the magnified view of the measurement in the low temperature range showing the positions of turn-on temperature T* as pointed by the arrows. In both, the case the T* is shifting right towards higher temperature with increasing magnetic fields.

Some published reports have emphasized that a shallow minimum at low temperatures with the resistivity increasing slightly towards lower temperatures can be perceived due to the weak localization or electronic correlations especially in amorphous and highly disordered systems[39,40]. The weak localization is observed due to coherent backscattering of conduction electrons induced by the increased disorder at low temperature[40] and contributes to negative magnetoresistance. However in our case, the observed magnetoresistance (MR) is positive (see details in magnetoresistance section). The enhanced electronic correlations may lead to reduction of the charge carrier density, consequently increasing resistivity, at the Fermi level by renormalization[39], which is not the case here as we have seen increased carrier concentration at low temperature (discussed in Hall measurements). Most importantly, weak localization and electronic correlation contributions are additive to the residual resistivity leading to an upturn in zero magnetic field $\rho_{xx}$ and that should vanish with application of higher magnetic fields. We have observed no upturn in zero field, but the appearance of an upturn under applied field. Thus, we rule out these possibilities. Turn-on or metal-semiconductor-like crossover behavior is an important feature observed in many topological materials in presence of strong magnetic fields. Ultra-thin flakes of WTe$_2$ as well as many other WSMs such as ZrSiS, and NbP[36–38,41], and unconventional semimetals like LaBi, and LaSb[35,42] have shown such phenomena. In these materials, the origin of such turn-on behavior has been primarily attributed to the gap opening at the band-touching points in presence of a strong magnetic field[36] and secondly to the competition between dominating scattering mechanisms that could occur at low temperatures[38]. LaAlGe being a topological semimetal[8], we strongly believe the former is the dominant reason while the later cannot be discarded. However, the origin of this behavior is still a subject of study and need further research.

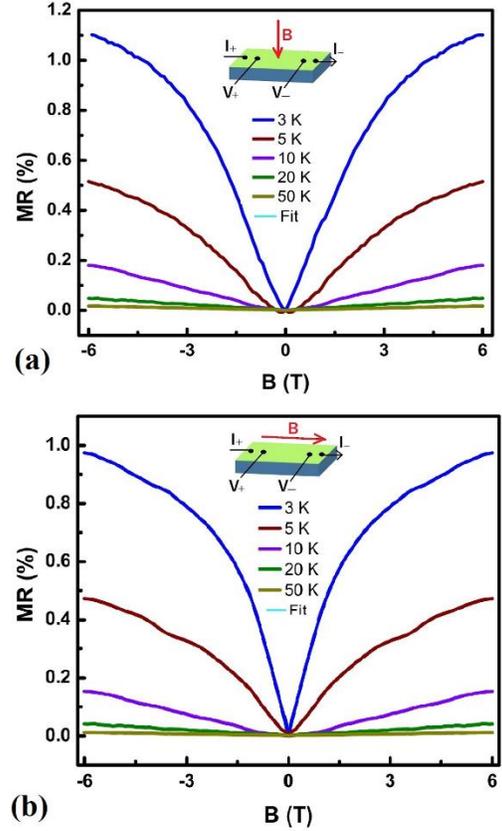

**Figure 5:** Magnetoresistance measurements at different temperatures. (a) Magnetoresistance as a function of magnetic field applied perpendicular to the direction of current flow. (b) Magnetoresistance as a function of magnetic field applied parallel to the direction of current flow. The schematics for the direction of magnetic field and current during the measurements is shown in both cases.

Figures 5 (a) and (b) display the evolution of perpendicular magnetoresistances (magnetic field applied perpendicular to the direction of current flow) and parallel magnetoresistances (magnetic field applied along the direction of the current flow) with temperature, measured in the field range ±6 T. Expressed as a percentage, magnetoresistance is given by the formula: $\frac{\rho_{xx}(B)-\rho_{xx}(0)}{\rho_{xx}(0)} \times 100\%$, where $\rho_{xx}(B)$ and $\rho_{xx}(0)$ are the resistivities at magnetic field $B$ $(= \mu_0 H)$ and zero field respectively. We observed positive MR for both parallel and perpendicular magnetic fields at low temperature regime from 3 to 20 K and it is quite interesting that the MR has small quadratic contribution at low field, but appears quasilinear that tends to saturate at higher magnetic fields. With the maximum magnetoresistance of ~1.10(1)% at 3 K in a field of 7 T, it is also observed that the MR considerably decreases with increasing temperature, while the shape of the curves at like temperatures appears almost identical for MR measured in both cases. There is no significant MR observed at 50 K and it almost disappears in the high temperature regime (not shown here), which might be due to the strong phonon-electron scattering. Bedoya-Pinto et al. [43] has shown similar positive MR in thin films of NbP in both for in-plane (B // I) and out-of plane (B ⊥ I) magnetic fields, suggesting the absence of chiral anomaly, due to contribution of nontopological bands located in momentum space. The effect of chiral charge reduces away from the Weyl points and the chiral anomaly is highly sensitive to the position of Fermi energy. This hints a possible shift of type II Weyl nodes from the Fermi level in thin film LaAlGe, which in contrast, is observed exactly at the Fermi level in bulk single crystal [8]. Additional study will be necessary for a more exhaustive analysis.

Hall resistivity $\rho_{xy}$ measurements taken at various temperatures starting from room temperature (300 K) down to 3 K and in a magnetic field up to 4 T are shown in Figure 6 (a). The $\rho_{xy}$ measured in between $300 - 5$ K is linear as a function of magnetic field, which suggests that the Fermi surface probably consists of electronic bands that are compensated, with electrons and holes having comparable overall density and mobilities[8,25,44]. Hence we analyzed and interpreted the $\rho_{xy}$ data using linear model (consequence of Lorentz force) $\rho_{xy} = R_0 B$, to obtain the Hall effect coefficient $R_0 = -1/(ne)$ where, $n$ is the effective charge carrier density, and $e$ the electronic charge. At 300 K, from the linear fit to the Hall resistivity data we obtained $R_0 = -6.45(2) \times 10^{-8}$ $\Omega$ $cm$ $T^{-1}$, where negative sign indicates the charge transport is dominated by the electrons and their corresponding effective carrier density $n = 9.68(1) \times 10^{21} cm^{-3}$. Similarly, at 10 K the average charge carrier concentration is $8.23(1) \times 10^{21} cm^{-3}$, which is slightly higher than the one obtained in the single crystals LaAlGe $(1.08(1) \times 10^{21} cm^{-3})$ [28]. At 3 K, the Hall resistivity curve displays a sign change from positive to negative on increasing the magnetic field above 1 T, which suggests that electronic bands at the Fermi surface are probably not compensated and both the electrons and holes contribute to the electrical transport properties at this temperature. Such a nonlinear Hall effect is expected in Weyl semimetals and has been recently reported in NbIrTe$_4$[45]. The fit to the linear part above 1 T is shown in Figure 6 (b). Figure 6 (c) summarizes the average carrier concentration calculated at different temperatures including at 3 K which confirms the semimetallic nature of our thin film samples. Noticeably, the average charge carrier concentration above 5 K shows gradual increment with rising temperature.

Overall, the charge carrier density is nearly stable showing weak temperature dependence, as in the unconventional semimetal LaBi[42] and CeAlGe[29].

Finally, the Hall conductivity was calculated using the formula $\sigma_{xy} = -\rho_{xy}/(\rho_{xy}^2 + \rho_{xx}^2)$ where the symbols have usual meanings. The magnetic field dependence of the Hall conductivity is displayed in Figure 6 (d). At 3 K, the Hall conductivity initially increases with the increasing magnetic field until ~ 1 T where it reaches a peak and thereafter decreases gradually with increasing field strength. The peak value of Hall conductivity observed is 43(1) S cm$^{-1}$ calculated by drawing a tangent to the Hall conductivity at the position of the peak. At temperatures higher than 3 K, we observed a linear Hall conductivity that remains unsaturated through 4 T.

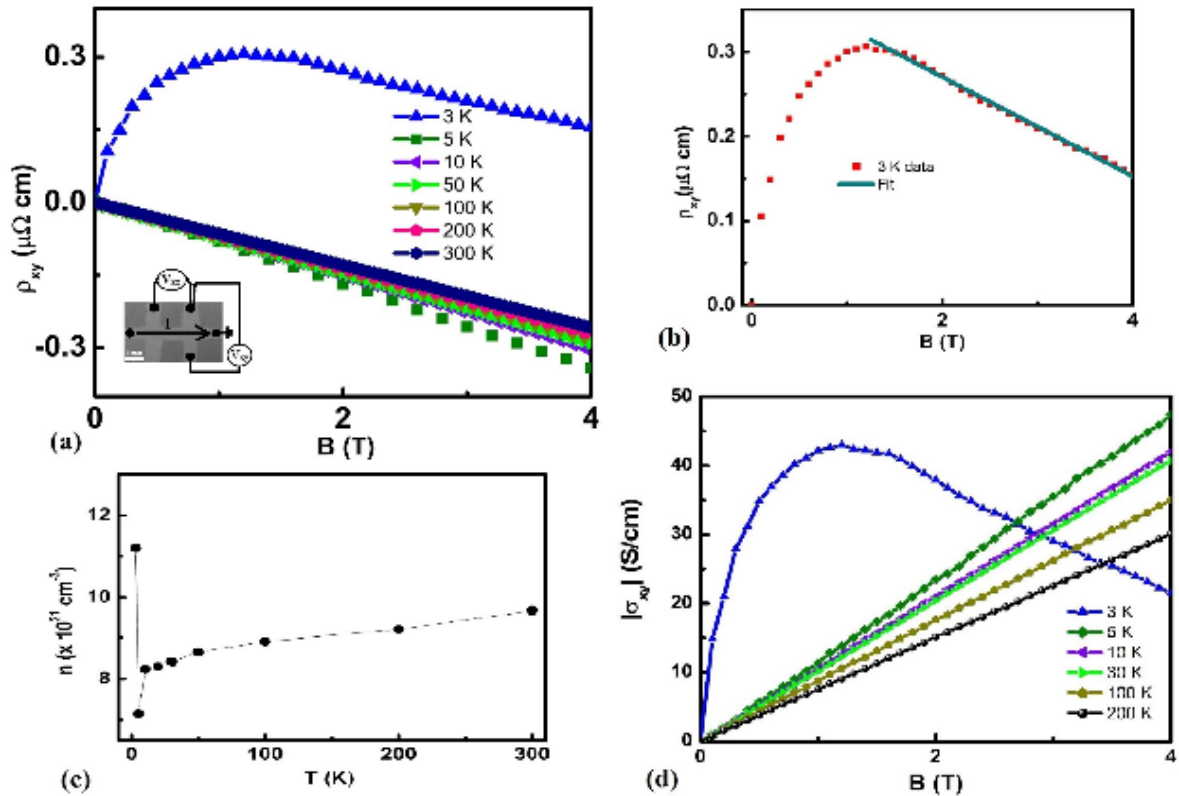

**Figure 6:** Hall Measurements: (a) Hall resistivity curves measured in between 3 and 300 K. Inset in the lower left corner show the Hall device used for the measurements. (b) Data at 3 K shows sign change near 1.5 T fields where higher field data is fitted linear. (c) Average carrier concentration calculated at corresponding temperatures. (d) Hall conductivity measurements taken in between 3 and 200 K.

## IV. SUMMARY AND CONCLUSIONS

In summary, we have grown, structurally characterized, and investigated the electrical and magnetotransport properties of ~ 50 nm thick LaAlGe films. Structural and morphological characterization using XRD, SEM, and AFM techniques suggest continuous and crystalline LaAlGe films have been grown with uniform surface topography. At low temperature regime, we observed a noticeable increase in longitudinal resistivity passing through field-dependent minima most likely due to a gap opening at the band touching points in presence of magnetic fields. We also observed a positive magnetoresistance, which tends to saturate at higher fields and eventually is suppressed beyond 50 K. The Hall measurements confirm the semimetallic conduction dominated by electrons, with an average charge carrier density of $9.68(1) \times 10^{21}$ $cm^{-3}$ at room temperature. These results are interesting as MBE-based atomically engineered heterostructures of WSM, and other materials can serve as an interesting platform to study the interfacial effects.

## ACKNOWLEDGEMENTS

The authors want to thank the Vitreous State Laboratory for providing the facilities required for conducting this research and for its financial support.

## DATA AVAILABILITY

The supporting data generated during this research are available from the corresponding author upon reasonable request.